\documentclass{article}

\usepackage[final]{DLI_2023}

\usepackage{comment}
\usepackage[utf8]{inputenc} 
\usepackage[T1]{fontenc}    
\usepackage{hyperref}       
\usepackage{url}            
\usepackage{booktabs}       
\usepackage{amsfonts}       
\usepackage{nicefrac}       
\usepackage{microtype}      
\usepackage{xcolor}         
\usepackage{graphicx}
\title{AI Recommendation System for Enhanced Customer Experience: A Novel Image-to-Text Method}

\author{%
Mohamaed Foued Ayedi\\
The University of Carthage\\
Higher School of Communications of Tunis, Tunisia\\
\texttt{mohamedfoued.ayedi@supcom.tn} \\
  \And
Hiba Ben Salem\\
The University of Carthage\\
Higher School of Communications of Tunis, Tunisia\\
\texttt{hiba.bensalem@supcom.tn} \\
  \And
Soulaimen Hammami\\
Stile.ai\\
\texttt{soulaimen.hammami@stile.ai} \\
 \And
Ahmed Ben Said\\
Department of Computer Science and Engineering\\
Qatar University, Doha, Qatar\\
\texttt{abensaid@qu.edu.qa} \\
\And
Rateb Jabbar\\
KINDI Center for Computing Research, College of Engineering\\
Qatar University, Doha, Qatar\\
\texttt{rateb.jabbar@qu.edu.qa} \\
\And
Achraf CHabbouh\\
Stile.ai\\
\texttt{achraf.chabbouh@stile.ai} \\
}

\begin{document}
\maketitle

\begin{abstract}
Existing fashion recommendation systems encounter difficulties in using visual data for accurate and personalized recommendations. This research describes an innovative end-to-end pipeline that uses artificial intelligence to provide fine-grained visual interpretation for fashion recommendations. When customers upload images of desired products or outfits, the system automatically generates meaningful descriptions emphasizing stylistic elements. These captions guide retrieval from a global fashion product catalog to offer similar alternatives that fit the visual characteristics of the original image. On a dataset of over 100,000 categorized fashion photos, the pipeline was trained and evaluated. The F1-score for the object detection model was 0.97, exhibiting exact fashion object recognition capabilities optimized for recommendation. This visually-aware system represents a key advancement in customer engagement through personalized fashion recommendations.
\end{abstract}

\section{Introduction}
The rapid expansion of E-commerce has transformed online retail, with global sales reaching \$4.28 trillion in 2020~\cite{Company2021}. This growth is attributed to increased internet access, improved logistics, shifting consumer behaviors, and greater product variety online ~\cite{Smith2019}. However, the vast assortment has complicated purchase decisions for fashion shoppers. Inaccurate recommendations lead to dissatisfaction and lower conversion rates ~\cite{Park&al2017}. Therefore, precise, personalized recommendation systems are critical for fashion E-retailers. Traditional systems relying on metadata, static images, and collaborative filtering struggle to capture nuanced aesthetics ~\cite{Gajic&Balzano2018}. The evolution in Artificial Intelligence, especially in areas like machine learning and natural language processing, has opened up new possibilities to address many of the current challenges in e-commerce, enabling more accurate, efficient, and user-centric solutions ~\cite{said2019probabilistic,said2018deep,baccour2023zero,baccour2021intelligent,baccour2022pervasive,abulibdeh2022impact,jabbar2018applied,moulahi2023privacy,jabbar2022recent}.
Recent research underscores the need for visually-aware, personalized recommendations in fashion  ~\cite{Jagadeesh&al2014} ~\cite{Corbière&al2017}. Advances in computer vision and deep learning enable richer representations of fashion images and individual preferences. Building on precedents like Google Image Search and Amazon's system \cite{Tashjian&al2019}. This study proposes a novel end-to-end fashion product recommendation system leveraging computer vision and deep learning models. The system incorporates object detection using YOLO-v8 ~\cite{Redmon&Farhadi2018} and product classification through FashionClip ~\cite{Gao&al2021} to understand customers' visual preferences from product images. Descriptive captions are generated using BLIP ~\cite{Li&al2021}, an image captioning technique. Retrieval and recommendations are enabled through efficient search over a product catalog scraped from retailers globally using OpenSearch ~\cite{Gormley&Tong2015}.

\section{Methodology}
In this section, we present the details of our proposed recommendation system.
Figure 1 depicts the details of the proposed recommendation system. The process starts with an outfit image as input to a fine-tuned object detection model, enabling us to isolate each distinct product. The cropped images are then passed through a zero-shot classification model to generate the precise label of the product. Moving forward, the product image, along with its label, is passed to an image captioning model to generate detailed descriptions of the item’s pattern and color.

\begin{figure}[h!]
  \centering
  \includegraphics[width=11cm, height=7cm]{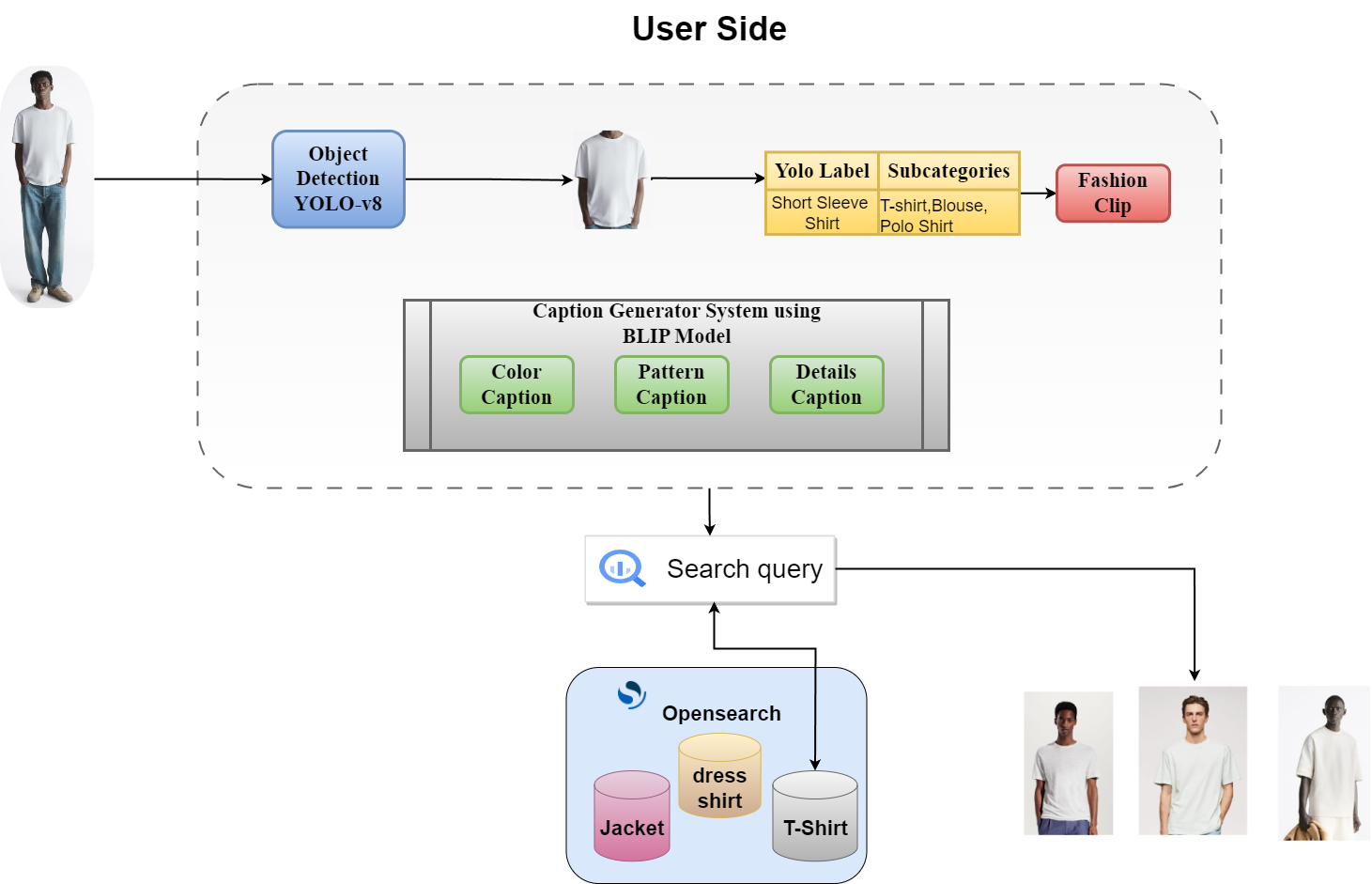}
  \caption{Overview of the proposed architecture.}
  \label{fig:label}
\end{figure}

To manage our data, we employed OpenSearch. In this setup, product labels serve as indexes for our search clusters. The final step in this procedure involves locating similar items.
In the following sections, we will describe each part of the system in detail.

\subsection{Data preparation}
We started with curating and balancing a dataset of 111,824 images sourced from web scraping. It is essential to have a balanced dataset when training object detection models like YOLO-v8 to avoid any class bias.

\subsection{Training and evaluation process}
The training procedure consisted of fine-tuning YOLO-v8 model for 100 epochs  on NVIDIA RTX 4090  for approximately 35 hours.
Our dataset was divided, with 90\% dedicated to training and the remaining 10\% used for testing and validation. We monitored three loss metrics: box loss, class loss, and defocus loss. Box loss relates to errors in bounding box prediction, class loss deals with errors in class prediction, and defocus loss considers errors in focus prediction.

\subsection{Object detection}
After training, we used our model to detect objects in input images. This allowed us to categorize and crop images of products into distinct items, such as long sleeve tops, short sleeve tops, long sleeve outerwear, trousers, and shorts. The main goal of this step was to analyze each part of an outfit independently.

\subsection{Products Classification}

For precise classification of fashion products, subcategories were created for each YOLO-recognized class. FashionClip ~\cite{6}, a CLIP model \cite{7} adaptation, was then used. It was trained on a dataset of over 700,000 image-text pairs from Farfetch ~\cite{ParamAggarwal2020}, a well-known luxury fashion retailer, and excels in ``zero-shot generalization``. Each cropped image, along with its YOLO-v8 label and subcategory, is processed by FashionClip, which generates the most fitting label within the subcategory by comparing image and labels embeddings and selecting the label with the highest cosine similarity score.

\subsection{Caption Generation}

In this step, we employ cropped images, coupled with their new labels, as input to an image captioning model to generate precise descriptions. We opted for BLIP~\cite{8}, a bootstrapping Language-Image Pretraining for Unified Vision, for its precise alignment between visual and textual content. BLIP employs the Vision Transformer (ViT) approach ~\cite{9}, to segment the input image into patches and encodes them as a sequence of embeddings. The BLIP components, Image-Text Contrastive Loss (ITC), Image-Text Matching Loss (ITM), and Language Modeling Loss (LM) link, evaluate, and generate text descriptions respectively. 
To slightly augment the output from BLIP, we used a prompt-based approach. At the start of each caption, prompts such as "this {label} features" are included, directing the generation process toward specific outfit aspects.

\subsection{Product Similarity Recommendations}
We enhanced our recommendation system by implementing data categorization across OpenSearch clusters. By using the product label from Fashion Clip (FC) and the captions generated using BLIP, we employed the match query for full-text searches. This approach considers word proximity and generates ranked recommendations based on matching scores. The integration of Okapi BM25, a powerful ranking function, within OpenSearch resulted in performance improvements, delivering faster and more precise suggestions ~\cite{10}.

\section{ Experiments }

We rigorously evaluate our fashion product recommendation system through a series of quantitative and qualitative experiments.
The experiments are conducted on dedicated test datasets, and we provided category-wise evaluations to gain comprehensive insights into our system's capabilities.

\subsection{Object Detection Results (YOLO-v8)}

As illustrated in Figure 2, the YOLO-v8 model achieved impressive results in object detection, with an average of 0.97 accuracy across five categories. Figure 3 further exhibits the Precision-Recall Curve for our model. However, occasional challenges were encountered when differentiating between long sleeve tops and outerwear.

\begin{figure}[h!]
  \centering
  \includegraphics[width=6cm]{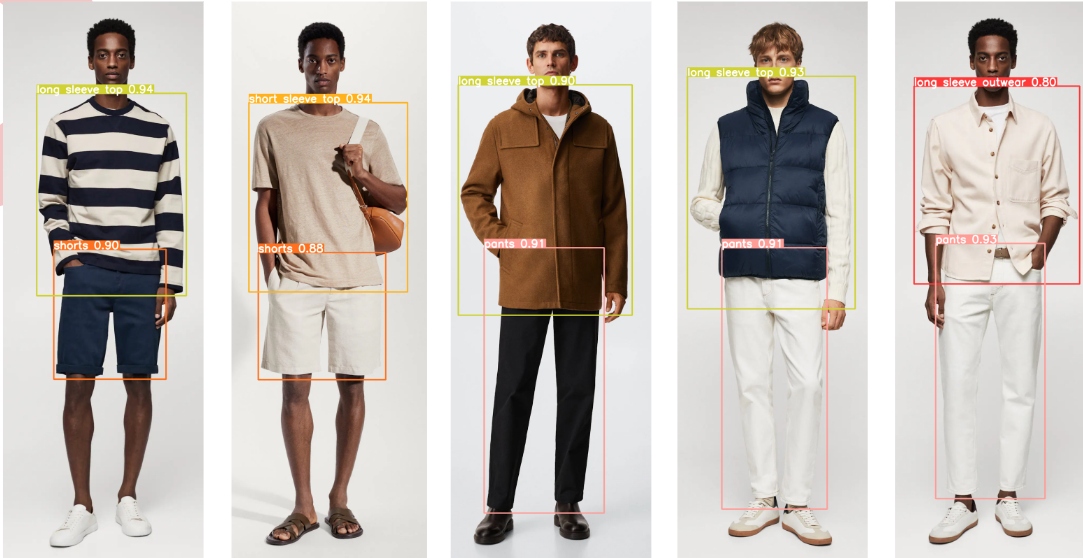}
  \caption{Object Detection Results with YOLO-v8 on Men Outfits. }
  \label{fig:label}
\end{figure}

\begin{figure} [h!]
  \centering
  \includegraphics[width=8cm]{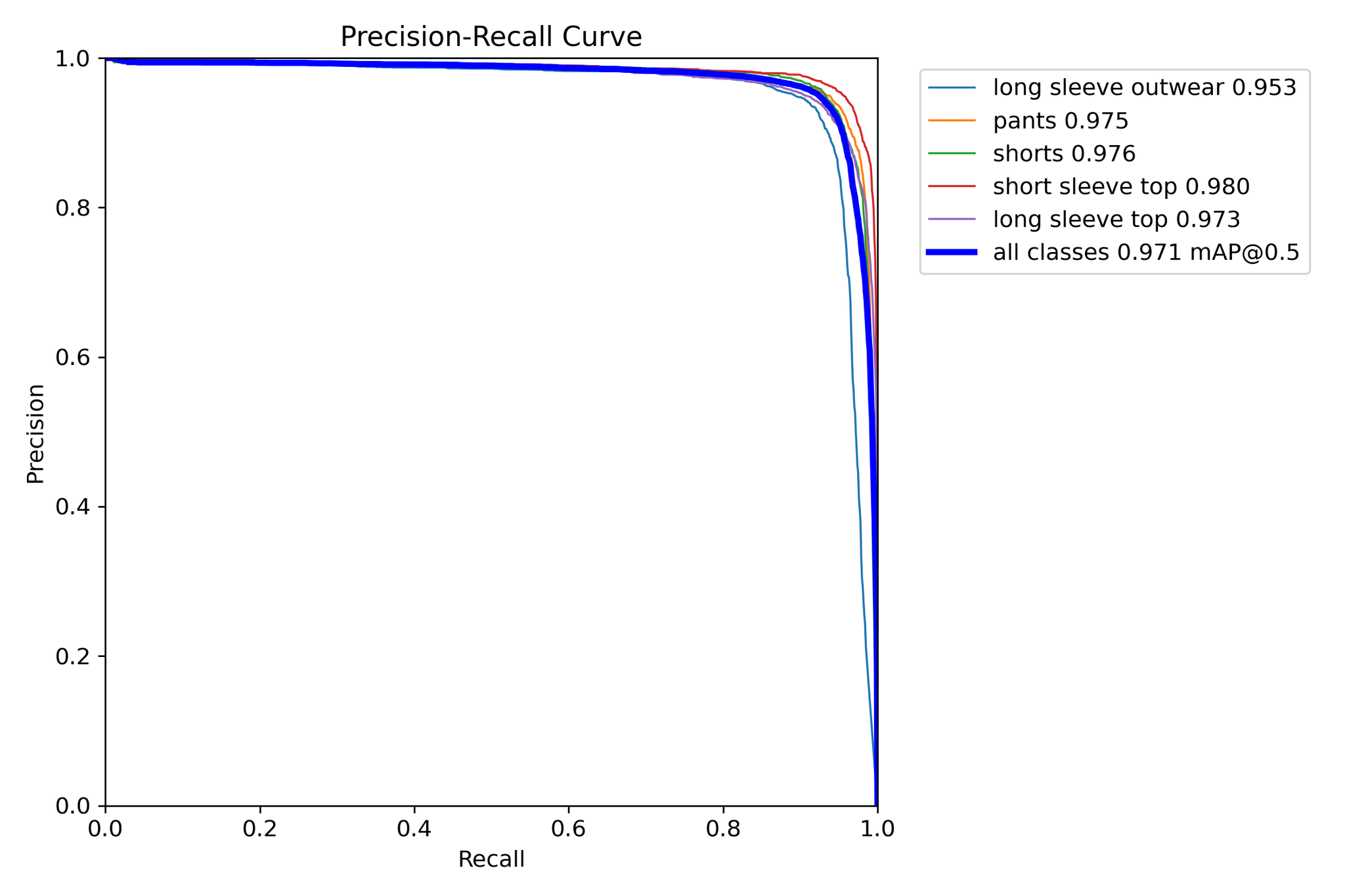}
  \caption{Precision-Recall Curve for Object Detection}
  \label{fig:label}
\end{figure}

\subsection{End-to-End Evaluation: image containing examples of generated product similarities }
\begin{figure}[h!]
  \centering
  \includegraphics[width=12cm, scale=1]{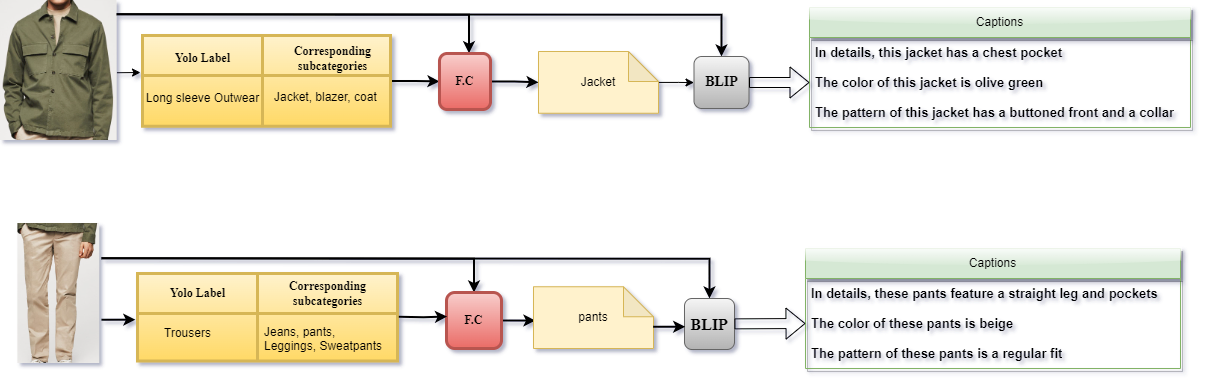}
  \caption{Captions generation results}
  \label{fig:label}
\end{figure}

BLIP, the caption generation model, excelled in providing intricate details about fashion products, including colors, patterns, and designs. This capability enhanced the quality of our products recommendations. Additionally, the fashion clip's accurate classification labels were instrumental in creating effective search clusters. Figure 4 shows an example of FashionClip and BLIP results.

Overall, our system consistently delivers strong results for fashion products, and Figure 5 exemplifies some similarity results, affirming the system's effectiveness in providing relevant fashion product recommendations.
\begin{figure}[h!]
  \centering
  \includegraphics[width=12cm]{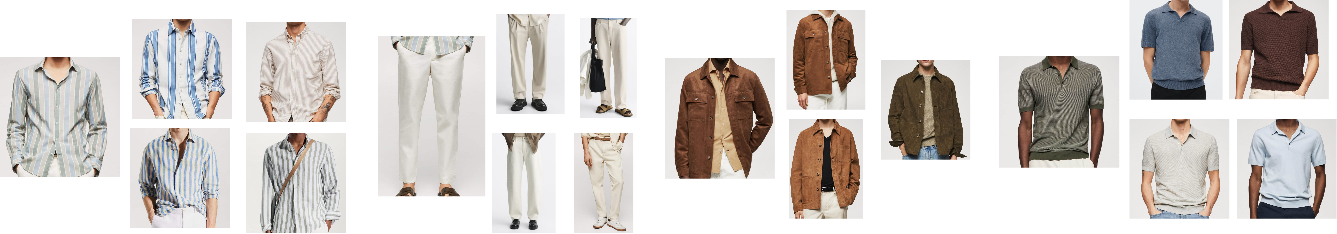}
  \caption{End-to-end results of the visual search system.}
  \label{fig:label}
\end{figure}
 For each query image, we present the top retrieved results.
\section{ Conclusion}
Our research has successfully engineered a novel fashion product recommendation system, with promising results in terms of accuracy. This system, utilizing a fine-tuned process of object detection, zero-shot classification, and image captioning, has the potential to transform online fashion retail.
Nevertheless, we have encountered certain challenges during this process. The success of our system depends on the quality of the input image and the accuracy of each model component. If any component fails to identify or describe an item correctly, it could have a significant impact on the results.
Going forward, our future work holds great potential. We aim to expand our system's capabilities to suggest not only similar items but also complementary outfit items. This would enrich the user shopping experience by providing comprehensive outfit suggestions.

\bibliographystyle{ieeetr}

\end{document}